\documentclass[12pt]{revtex4}

\usepackage{amsmath}
\usepackage{amsfonts}
\usepackage{amssymb}
\usepackage{graphicx}
\usepackage{pdfpages}

\newcommand{\ud}{\mathrm{d}}

\newcommand{\uD}{\mathrm{D}}

\begin{document}

\title{
Variational formulation of compressible hydrodynamics  
in curved spacetime and symmetry of stress tensor
}
\author{T.\ Koide}
\email{tomoikoide@gmail.com,koide@if.ufrj.br}
\affiliation{Instituto de F\'{\i}sica, Universidade Federal do Rio de Janeiro, C.P. 68528,
21941-972, Rio de Janeiro, RJ, Brazil}
\author{T.\ Kodama}
\email{kodama.takeshi@gmail.com,tkodama@if.ufrj.br}
\affiliation{Instituto de F\'{\i}sica, Universidade Federal do Rio de Janeiro, C.P. 68528,
21941-972, Rio de Janeiro, RJ, Brazil}
\affiliation{Instituto de F\'{\i}sica, Universidade Federal Fluminense, 24210-346,
Niter\'{o}i, RJ, Brazil}

\begin{abstract}
Hydrodynamics of the non-relativistic compressible fluid  
in the curved spacetime is derived using the generalized framework of the stochastic variational method (SVM) for continuum medium.
The fluid-stress tensor of the resultant equation becomes asymmetric for the exchange of the indices, differently from the standard Euclidean one. 
Its incompressible limit suggests that the viscous term should be represented with the Bochner Laplacian. 
Moreover the modified Navier-Stokes-Fourier (NSF) equation proposed by Brenner can be considered even in the curved spacetime. 
To confirm the compatibility with the symmetry principle, SVM is 
applied to the gauge-invariant Lagrangian of a charged compressible fluid and then the Lorentz force is reproduced as the interaction between 
the Abelian gauge fields and the viscous charged fluid.
\end{abstract}


%\begin{keywords}

\maketitle

\section{introduction}

The Navier-Stokes-Fourier (NSF) equation is used to describe non-relativistic behaviors of Newtonian fluids in the Euclidean spacetime. 
The fundamental requirements for the derivation are continuum medium picture, local thermal equilibrium and conservation laws.
The second requirement is, however, not justifiable in a rigorous manner and hence it is not obvious whether the coarse-grained dynamics of continuum medium 
can always be described in the analogous form of the NSF equation. 
Indeed, this is a common problem in the applications of hydrodynamical descriptions. For example, hydrodynamical models have successfully been applied to describe dynamics in relativistic heavy-ion collisions, 
where the realization of the local thermal equilibrium is highly nontrivial \cite{hydroreview}.

In this paper, we focus on non-relativistic hydrodynamics in curved spacetime systems. 
We accept all the aforementioned fundamental requirements.
A direct approach to find such a theory might be to reexpress the NSF equation in the Euclidean system with a general metric, 
but unfortunately such an approach is known to be controversial.
In the Euclidean spacetime, the NSF equation is described in the Cartesian coordinates. 
Then the local angular momentum conservation leads to the symmetry for the exchange of the indices of the fluid-stress tensor 
if there is no additional internal structure of fluid \cite{koidespinhydro}. 
This argument is however not applicable in the case of general coordinate systems, because 
the generalized momentum represents not only the linear momentum but also the angular momentum.
Thus the above symmetry of the stress tensor does not lead to the angular momentum conservation and vice versa.
That is, the standard requirement of hydrodynamics is not necessarily applicable to the derivation in the curved spacetime.
In addition, the physical origin of fluid viscosity can be affected by curved geometries.
Therefore 
the simple covariant reexpression of the standard NSF equation may not give an appropriate equation.
The behaviors of such a fluid have been studied exclusively for the incompressible fluid on spatial Riemannian manifolds (See however Refs.\ \cite{bella,koba,li}). 
In these studies, the required equations are obtained by replacing the Laplacian in the viscous term of the standard NSF equation with the corresponding operator on the manifold.
Such a replacement is not unique and there are at least three different proposals: the Hodge Laplacian, the Bochner Laplacian and the Laplacian defined with the deformation tensor.
See Refs.\ \cite{ebin,avez,carve,din,il,khe,lich2016,mazz,mitrea,naga,pierf,taylor,zhang,temam,chan,samavaki,jankuhn,fang,bella,koba,yasue-ns,cruzeiro,li} and references therein.

Another formulation of hydrodynamics is variational approaches.
In this paper, we consider the stochastic generalization of the variational principle, called stochastic variational method (SVM). 
There are many works to derive the Euclidean NSF equation in the different formulations of SVM \cite{inoue,nakagomi,marra,gomes,eyink,del,nov}. 
See also Refs.\ \cite{marner,holm} as related papers.
The applications to the spatial Riemannian manifolds are discussed in Refs.\ \cite{yasue-ns,cruzeiro}. 
Note that all works mentioned here consider only the incompressible fluid.

The purpose of the present paper is to derive the compressible hydrodynamical equation 
of the non-relativistic fluid in the curved spacetime.
In Refs.\ \cite{koide12,koide18}, the present authors extended one of  SVM's which was formulated by Yasue \cite{yasue}
to the continuum medium in the Euclidean spacetime, and derived the compressible NSF equation.
On the other hand, the generalization of the formulation of SVM to the curved spacetime is developed separately in Ref.\ 
\cite{koide19}.
We combine these approaches to derive our hydrodynamical equations.
Our result shows that the symmetry of the derived fluid-stress tensor is violated due to the Ricci tensor.
Taking the incompressible limit of the derived equation, we further observe that the Laplacian in the viscous term should be replaced with the Bochner Laplacian.
In addition, the modified NSF equation proposed by Brenner can be considered even in the curved spacetime. 
One of the advantages of the variational formulation is its compatibility with the symmetry principle. 
As an example, we discuss the Abelian gauge interaction of a charged compressible fluid using the gauge-invariant Lagrangian, reproducing the Lorentz force in curved systems.

This paper is organized as follows.
The standard variation to derive the Euler equation is reviewed in Sec.\ \ref{sec:euler}.
The application of SVM to derive the hydrodynamical equation in the curved spacetime is studied in Sec.\ \ref{sec:nsf}.
The incompressible limit of the derived equation is discussed in Sec.\ \ref{sec:incom}.
There, Brenner's bivelocity hydrodynamics in the curved spacetime is investigated.
The comparisons with other results are investigated in Sec.\ \ref{sec:compa}.
In Sec.\ \ref{sec:gauge}, the gauge interaction of the charged compressible fluid is considered. 
Section \ref{sec:con} is devoted to the concluding remarks.

\section{Euler equation in curved spacetime} \label{sec:euler}

In this section, we discuss the derivation of the Euler equation which describes the ideal fluid 
using the standard variational method for the continuum medium. 
The variation of the internal energy term here is utilized in the SVM calculation of the next section.
As related references, see Refs.\ \cite{arnold,hugo}.

For the curved spacetime characterized by the differentiable metric $g_{\mu\nu}$, it is possible to establish a local Minkowskian coordinate with the metric $\eta_{ab}$ 
($= {\rm diag}( -1,1,1,\cdots )$ for ($D+1$)-dimensional spacetime). 
The spacetime coordinates with $g_{\mu\nu}$ is denoted by $x^{\mu}$ while those with $\eta_{ab}$ by $y^{a}$. 
The Greek indices $\alpha,\beta,\cdots$ are used to label the general coordinate.
The Latin indices $a,b,\cdots$ are for the local Minkowskian coordinate whereas $i,j,k,l,m,n$ are reserved to denote the spatial components of $x^\mu$.
The Einstein notation of the summation is used.

In the following calculation, we use the tetrads of the vielbein \cite{synge},  
\begin{eqnarray}
\underline{e}^\mu_a (x) = \left. \frac{\partial x^\mu}{\partial y^{a}} \right|_{x }\, , \quad
\overline{e}^{a}_\mu (x) = \left. \frac{\partial y^a}{\partial x^{\mu}} \right|_{x}\, , 
\end{eqnarray}
satisfying 
\begin{eqnarray}
g_{\mu\nu}\underline{e}^{\mu}_a \underline{e}^{\nu}_b = \eta_{ab}\, ,  \quad
\eta_{ab}\overline{e}^{a}_\mu \overline{e}^{b}_\nu = g_{\mu\nu}\, .
\end{eqnarray}

A non-relativistic system is considered and the time component is always given by 
\begin{eqnarray}
\ud x^0 = c \ud t\, ,  \label{eqn:cond1}
\end{eqnarray}
where $c$ is the speed of light.
We study a limited case of the metric tensor satisfying
\begin{eqnarray}
g_{0i} = 0\, , \quad \partial_i g_{00} = 0\, .\label{eqn:cond2}
\end{eqnarray}
The former indicates that there is no mixture of the spatial and temporal components and the latter that the time scale is spatially homogeneous.

For the formulation of SVM, we introduce the Lagrange coordinate system: 
we denote the trajectory of a fluid element by 
$x^\mu_t (\xi)$ where the initial spacetime point of the fluid element is given by $\xi$. 
The initial distribution of the fluid element is denoted by $\rho_{(0)}(\xi)$. 
The total mass of the fluid is then given by $\int \ud V_\xi \rho_{(0)}(\xi)$ 
with the volume element $\ud V_\xi$.
The scalar mass distribution in Eulerian coordinates is defined by
\begin{eqnarray}
\rho(x) = \frac{1}{\sqrt{-g}}\int c\ud t \int \ud V_\xi \, \rho_{(0)}(\xi) \delta^{(D)} (x^\mu - x^\mu_t (\xi))\, ,
\end{eqnarray}
where $ \delta^{(D)} (x^\mu - x^\mu_t (\xi))$ is the product of the Dirac delta function for spatial $D$-dimensional system.
 
By generalizing the fluid Lagrangian in the Euclidean spacetime \cite{koide18},
we have the corresponding Lagrangian in the curved spacetime as  
\begin{eqnarray}
L 
&=& 
\int \ud V_\xi \, \rho_{(0)} (\xi) \left[
\frac{1}{2} \frac{\ud y^{a}(x_t)}{\ud t} \eta_{ab} \frac{\ud y^{b}(x_t)}{\ud t} 
- \frac{\varepsilon ( \rho_{(0)}(\xi) /J (\partial x_t) )}{ \rho_{(0)}(\xi) /J (\partial x_t)  }
\right]\, , \label{eqn:lag-euler-lag2}
\end{eqnarray}
where $\varepsilon(\rho)$ is the scalar distribution of $\rho$ and 
the Jacobian is defined by 
\begin{eqnarray}
J (\partial x_t) = \left|
\frac{\partial y(x_t)}{\partial y(\xi)}
\right|\, ,
\end{eqnarray}
and 
\begin{eqnarray}
\ud y^{a}(x) = \ud x^\mu \bar{e}^{a}_\mu (x) \, .
\end{eqnarray}
As shown soon below, $\varepsilon(\rho)$ represents the internal energy of the fluid which locally satisfies the thermodynamical relation (local thermal equilibrium). 
In the above definition of the Lagrangian, we expressed the kinetic term (the first term) in terms of the variables of the local Minkowskian coordinates. 
It is because we have to consider the changes of not only $x^\mu_t$ but also $\overline{e}^{a}_\mu(x_t)$. 
Note that the kinetic term contains the contribution from the time components of $y^0$, 
but it is just a constant term and does not contribute to the variation because of Eqs.\ (\ref{eqn:cond1}) and (\ref{eqn:cond2}).

Let us consider the variation only for the spatial components of $x^\mu_t (\xi)$, defined by
\begin{eqnarray}
x^{\mu}_t (\xi) \longrightarrow x^{\mu}_t (\xi) + \delta f^{\mu}(\vec{x}_t,t)\, , \label{eqn:smoothvari1}
\end{eqnarray}
where $\delta f^{0} (\vec{x},t) = 0$ and 
\begin{eqnarray}
\delta f^{\mu} (\vec{x},t_i) = \delta f^{\mu} (\vec{x},t_f) = 0\, .\label{eqn:smoothvari2}
\end{eqnarray}
We introduced $\vec{x}$ to denote all spatial components of $x^{\mu}$.
Note that $\delta f^{\mu} (\vec{x},t)$  is infinitesimal and smooth for the changes of $x^{i}$ and $t$.
Then the variation of the Lagrangian (\ref{eqn:lag-euler-lag2}) can be calculated as 
\begin{eqnarray}
&& 
\int \ud V_\xi \rho_{0}(\xi)
\left[
  g_{ij} \frac{\ud^2 x^{i}_t}{\ud t^2}\delta f^{j} (\vec{x}_t,t) - \frac{P}{\rho^2(x_t)}  \delta \rho (x_t)
\right]
= 0 \, , 
\end{eqnarray}
where the adiabatic pressure is defined through the thermodynamical relation as  
\begin{eqnarray}
P = - \frac{\ud}{\ud \rho^{-1} } \frac{\varepsilon}{\rho}\, . \label{eqn:press}
\end{eqnarray}
That is, we consider the barotropic fluid. 
See also Refs. \cite{koide12,koide18,hugo}.
We further use the variation of the scalar mass distribution which is calculated as
\begin{eqnarray}
\delta \rho(x_t) = -\frac{\rho^2 (x_t)}{\rho_{(0)}(\xi)} {A_{b}}^{a} (x_t)  \frac{\partial}{\partial y^a (\xi)} \delta f^\mu(\vec{x}_t,t)  \bar{e}^b_\mu (x_t) \, ,
\end{eqnarray}
where 
\begin{eqnarray}
{A_{b}}^{a} (x) = \frac{\partial J (\partial x)}{\partial (\partial y^a (x)/\partial y^b (\xi))}\, ,
\end{eqnarray}
is the cofactor of the Jacobian matrix element, which satisfies 
\begin{eqnarray}
 \frac{\partial}{\partial y^a (\xi)} {A_{b}}^{a} (x) = 0\, ,
\end{eqnarray}
and
\begin{eqnarray}
 {A_{b}}^{a} (x) \frac{\partial}{\partial y^a (\xi)} = J(\partial x) \frac{\partial}{\partial y^b (x)}\, .
\end{eqnarray}
We eventually find that the variation leads to the expected Euler equation,  
\begin{eqnarray}
\rho v^\mu \nabla_\mu v^{i} = - g^{i\mu} \nabla_{\mu} P\, ,
\end{eqnarray}
where $\nabla_\mu$ is the covariant derivative and the four velocity is defined by 
\begin{eqnarray}
v^\mu = (c, \ud \vec{x}_t/ \ud t ) \, . \label{eqn:vel1}
\end{eqnarray}
This four velocity behaves as a vector under general coordinate transformations satisfying Eq.\ (\ref{eqn:cond2}). 
Note that, in the above Lagrange picture, we implicitly assumed that there is no turbulence where the concept of  
the trajectories of the fluid elements become obscure.

\section{Viscous equation in curved spacetime}  \label{sec:nsf}

In the above derivation of the Euler equation, 
we assumed that the trajectory of the fluid element is smooth and the fluid velocity is given by its time derivative.
See Eq.\ (\ref{eqn:vel1}).
However, 
strictly speaking, thermal fluctuations affect the motion of the fluid element and then 
the corresponding trajectory can be non-differentiable. 
In SVM, this effect is taken into account by assuming the Brownian motion for the trajectory of the fluid element \cite{koide12,koide18}. 
In the following, the stochastic quantity corresponding to, say, $x$ is denoted by $\widehat{x}$.
See also Ref.\ \cite{koide19} for details of the formulation developed in this section.

\subsection{ Forward and Backward Stochastic Differential Equations}

Suppose that the evolution of the trajectory forward in time ($\ud t>0$) is described by the following forward stochastic differential equation (SDE), 
\begin{eqnarray}
\ud \widehat{x}^{\, i}_t =  {u}^{i}_{+} (\widehat{x}_t) \ud t + \sqrt{2\nu} \underline{e}^{i}_a (\widehat{x}_t) \circ_s \ud \widehat{W}^a_t\, ,\label{eqn:fsde}
\end{eqnarray}
where 
$u^{i}_+(x)$ is still unknown smooth velocity field and 
$\nu$ is a parameter to characterize the noise intensity which is associated with thermal fluctuations of microscopic constituent particles of the fluid.
The standard Wiener process $\widehat{W}^a_t$ has only spatial components ($\widehat{W}^{0}_t = 0$) which satisfy
\begin{eqnarray}
E[\ud \widehat{W}^a_t] = 0\, ,  \quad
E[(\ud \widehat{W}^a_t)(\ud \widehat{W}^b_{t^\prime})] = \ud t \, \delta^{a b} \delta_{t,t^\prime}\ \ \ (a,b \neq 0)\,  .  \label{eqn:w2}
\end{eqnarray}
The stochastic ensemble average is denoted by $E[\quad ]$.
In Eq.\ (\ref{eqn:fsde}), we introduced the Stratonovich definition of the product defined by 
\begin{eqnarray}
f(\widehat{x}_t) \circ_s \ud\widehat{W}^a_t = f(\widehat{x}_{t+ (\ud t/2)}) \ud \widehat{W}^a_t\, , \label{eqn:stra}
\end{eqnarray}
for an arbitrary smooth function $f(x)$.
Note that Leibniz's rule of differential for stochastic quantities is formally held when the Stratonovich definition is applied \cite{book:gardiner}. 
The purpose of SVM is to find the unknown smooth function $u^i_+(x)$ 
by the variational principle.

The change of the tetrad is determined by 
the Levi-Civita-Ito stochastic parallel transport \cite{ito}, 
\begin{eqnarray}
\ud \underline{e}^{\mu}_a (\widehat{x}_t) 
=
-\Gamma^\mu_{\nu \delta} (\widehat{x}_t) \underline{e}^\delta_a (\widehat{x}_t) \circ_s \ud \widehat{x}^\nu_t\, , \label{eqn:ev} 
\end{eqnarray}
where $\Gamma^\mu_{\nu \delta}(x)$ is the Christoffel symbol.
The length of the transported vector is conserved in this definition, while different choices can be considered. See the discussion in Sec.\ \ref{sec:compa}.
Both of Eqs.\ (\ref{eqn:fsde}) and (\ref{eqn:ev}) can be reexpressed in terms of the Ito definition of the product. 
See Appendix \ref{app:ito}.

In the variational principle, we should fix not only an initial condition but also a final condition. 
This implies that the forward SDE alone is not sufficient \cite{zam-review,koide-review}.  
We further introduce the backward SDE for $\ud t<0$ as 
\begin{eqnarray}
\ud \widehat{x}^{\, i}_t =  {u}^{i}_-(\widehat{x}_t) \ud t + \sqrt{2\nu} \underline{e}^{i}_a (\widehat{x}_t) \circ_s \ud \underline{\widehat{W}}^a_{\, t}\, ,
\end{eqnarray}
where ${u}^{i}_-(x)$ is another unknown function and $\underline{\widehat{W}}^a_{\, t}$ is another Wiener process 
which satisfies the same correlations as $\widehat{W}^{a}_t$ by replacing $\ud t$ by $|\ud t|$.

The backward SDE should correspond to the time-reversed process of the forward SDE.  
Thus there exists a condition associating $u^i_-(x)$ with ${u}^i_+(x)$. 
To find it, it should be noted that the scalar mass distribution in the previous section is replaced by the expectation value,
\begin{eqnarray}
\rho(x) = \frac{1}{\sqrt{-g}} \int c \ud t \int \ud V_\xi \, \rho_{(0)}(\xi) E[\delta^{(D)} (x^\mu - \widehat{x}^{\, \mu}_t (\xi))]\, .
\label{eqn:scalarmass-sto}
\end{eqnarray}
Substituting the solution of the forward SDE, 
the evolution equation of $\rho$, which is called the Fokker-Planck equation, is given by 
\begin{eqnarray}
\lefteqn{c\nabla_0 \rho (x)} && \nonumber \\
&=& -\nabla_j (u^{j}_+ (x)\rho(x)) + \nu \Delta_{LB} \rho (x)- \nu (\nabla_i \rho (x)) 
\left[
\Gamma^j_{j0}(x) g^{0i} (x)
+ \Gamma^{i}_{j0} (x)g^{0j} (x)
- g^{ij} (x)\Gamma^0_{0j}(x)
\right] \nonumber \\
&=& -\nabla_j (u^{j}_+ (x)\rho (x)) + \nu \Delta_{LB} \rho (x) \, ,\label{eqn:fpf}
\end{eqnarray}
where $\Delta_{LB} = g^{ij} \nabla_i \partial_j$ is the Laplace-Beltrami operator but the sum runs only for 
the spatial components $i,j=1,2,\cdots,D$ 
because of the correlation defined by Eq.\ (\ref{eqn:w2}).
To find this result, we used Ito's lemma (Ito formula) \cite{book:gardiner} and the condition given by Eq.\ (\ref{eqn:cond2}).
The above definition of $\rho$ should be modified when we consider the periodic variable such as angle, 
but the following results are not affected by such a modification. See Appendix\ \ref{app:periodic}.

On the other hand, substituting the solution of the backward SDE, we find another Fokker-Planck equation, 
\begin{eqnarray}
c\nabla_0 \rho (x)
&=& -\nabla_j (u^{j}_- (x) \rho (x) ) - \nu \Delta_{LB} \rho (x) \, .\label{eqn:fpb}
\end{eqnarray}
These two Fokker-Planck equations should be equivalent and thus we arrive at the following consistency condition, 
\begin{eqnarray}
u^{i}_+ (x)= u^{i}_- (x) + 2 \nu g^{i\mu}(x)\partial_\mu \ln \rho(x)\, . \label{eqn:cons-cond}
\end{eqnarray}
See also the discussion around Eq.\ (17)  in Ref.\ \cite{koide12}.
Then the two Fokker-Planck equations are reduced to the same equation of continuity, 
\begin{eqnarray}
\nabla_\mu (\rho(x) v^\mu(x)) = 0\, , \label{eqn:eoc}
\end{eqnarray}
with the four velocity defined by 
\begin{eqnarray}
v^\mu (x) 
= (c, \vec{v}(x)) \, , \quad
\vec{v}(x)
= \frac{\vec{u}_+ (x) + \vec{u}_-(x)}{2}\, .
\end{eqnarray}
This definition coincides with that of the Euler equation (\ref{eqn:vel1}) in the vanishing limit of the noise, $\nu\longrightarrow 0$.

To apply the variational principle, we need to construct an action which is a functional of a stochastic trajectory.
To define the kinetic term, we need to introduce the time differential of trajectory.
The stochastic particles follow zigzag paths 
and thus the standard definition of the particle velocity is not applicable. 
The possible time differentials are studied by Nelson \cite{nelson}: one is the mean forward derivative,
\begin{equation}
\uD_+  f(\widehat{x}_t)  = \lim_{\ud t \rightarrow 0+} E \left[  \frac{ f(\widehat{x}_{t + dt}) -
f(\widehat{x}_t)}{\ud t} \Big| \mathcal{P}_{t} \right]\,  ,
\end{equation}
and the other the mean backward derivative,
\begin{equation}
\uD_-  f(\widehat{x}_t)  = \lim_{\ud t \rightarrow0-} E \left[  \frac{ f(\widehat{x}_{t + dt}) -
f(\widehat{x}_t)}{\ud t} \Big| \mathcal{F}_{t} \right]\, .
\end{equation}
These expectation values are conditional averages, where $\mathcal{P}_{t}$ ($\mathcal{F}_{t}$) indicates to fix values of $\widehat{x}^{i}_{t^\prime}$ for
$t^{\prime}\le t~~(t^{\prime}\ge t)$. 
For the $\sigma$-algebra of all measurable events of $\widehat{x}_t$, $\mathcal{P}_{t}$ and $\mathcal{F}_{t}$
represent an increasing and a decreasing family of sub-$\sigma$-algebras, respectively. 
These derivatives are connected through the stochastic partial integration \cite{koide-review},
\begin{eqnarray}
\int^{b}_a \ud s E[\widehat{Y}_s \uD_+ \widehat{X}_s] = - \int^{b}_a \ud s E[\widehat{X}_s \uD_- \widehat{Y}_s] 
+ \int^b_a ds \frac{\ud}{\ud s} E[\widehat{X}_s \widehat{Y}_s]\, .\label{eqn:pif}
\end{eqnarray}

We suppose the existence of the background smooth curved spacetime and thus we can define the stochastic quantity $\widehat{y}^{a}_t \equiv y^{a}(\widehat{x}_t)$ at each spacetime point.
Then applying Ito's lemma, we find 
\begin{eqnarray}
\underline{e}^{i}_a  (\widehat{x}_t) \ud \widehat{y}^{\, a}_t &=& \underline{e}^{i}_a (\widehat{x}_t) \, \ud \widehat{x}^{\, \mu}_t \circ_s \overline{e}^{a}_\mu (\widehat{x}_t) \nonumber \\
&=& \underline{e}^{i}_a (\widehat{x}_t) \ud \widehat{x}^{\, \mu}_t \overline{e}^{a}_\mu (\widehat{x}_t) - \frac{1}{2}\overline{e}^{a}_j (\widehat{x}_t) \ud \widehat{x}^{\, j}_t 
\ud \underline{e}^{i}_a (\widehat{x}_t)
\nonumber \\
&=& 
\ud \widehat{x}^{\, i}_t \pm \nu \ud t g^{jk} (\widehat{x}_t) \Gamma^{i}_{jk} (\widehat{x}_t)\, , \label{eqn:edy}
\end{eqnarray}
where the sign $+$ $(-)$ represents the result using the forward (backward) SDE.
In the first line, the product of $\ud \widehat{x}^\mu_t$ and $ \overline{e}^{a}_\mu$ follows the Stratonovich definition.  
In this derivation, we used
\begin{eqnarray}
\begin{split}
&& \ud \overline{e}^{a}_\mu \ud \underline{e}^{i}_a = - 2\nu \ud t \Gamma^{i}_{j\delta} \Gamma^{\delta}_{k \mu} g^{jk}\, , \\
&& \underline{e}^\nu_{a}\circ_s \ud \overline{e}^{a}_\mu +  \overline{e}^{a}_\mu \circ_s \ud \underline{e}^\nu_{a} = \ud(\underline{e}^\nu_{a}\overline{e}^{a}_\mu) = 0\, .
\end{split}
\end{eqnarray}
Substituting Eq.\ (\ref{eqn:edy}) into the definitions of the mean forward and backward derivatives, we observe 
\begin{eqnarray}
\underline{e}^{i}_a  (\widehat{x}_t) \uD_{\pm} \widehat{y}^{a} = u^{i}_{\pm} (\widehat{x}_t)\, .
\end{eqnarray}
Here we used the result in Appendix \ref{app:ito}.

Similarly, for an arbitrary smooth vector function $A^{\mu}(x)$, we find 
 \begin{eqnarray}
\lefteqn{\underline{e}^{i}_a(\widehat{x}_t) \uD_{\pm} (A^{\mu} (\widehat{x}_t) \overline{e}^{a}_{\mu}(\widehat{x}_t)) } \nonumber \\
&&\hspace*{-0.5cm}=\underline{e}^{i}_a(\widehat{x}_t)  E\left[ \frac{\ud A^\mu(\widehat{x}_t)}{\ud t} \circ_s \overline{e}^{a}_\mu (\widehat{x}_t) + A^\mu (\widehat{x}_t)\circ_s \frac{\ud \overline{e}^{a}_\mu(\widehat{x}_t)}{\ud t}  \Big| \mathcal{ P}_t (\mathcal{F}_t) \right] \nonumber \\
&&\hspace*{-0.5cm}= \left( c \nabla_0
+ u^j_{\pm} (\widehat{x}_t)\nabla_j \pm  \nu g^{jk}(\widehat{x}_t) \nabla_j \nabla_k \right) 
A^i (\widehat{x}_t)\, . \label{eqn:ito-vector2}
\end{eqnarray}
Note that the derivative of 
the tetrad reproduces the corrections for the parallel transport in the curved geometry.

\subsection{Stochastic variational principle in curved spacetime}

Now, we introduce the stochastic Lagrangian which reproduces the 
classical one (\ref{eqn:lag-euler-lag2}) in the vanishing limit of  $\nu$.
Because of the two different time derivatives, the most general quadratic form of the kinetic term is given by 
\begin{eqnarray}
\lefteqn{\frac{1}{2}  \frac{\ud y^{a}(x_t)}{\ud t} \eta_{ab} \frac{\ud y^{b}(x_t)}{\ud t} } && \nonumber \\
&\longrightarrow &
B_+ \left\{ \frac{A_+}{2} (\uD_+ \widehat{y}^{\, a})\eta_{ab} (\uD_+ \widehat{y}^{\, b} )
+ \frac{A_- }{2} (\uD_- \widehat{y}^{\, a})\eta_{ab} (\uD_- \widehat{y}^{\, b}) \right\} 
 + \frac{B_-}{2} (\uD_+ \widehat{y}^{\, a}) \eta_{ab} (\uD_- \widehat{y}^{\, b})\, ,
\end{eqnarray} 
with 
\begin{eqnarray}
\begin{split}
A_\pm = \frac{1}{2} \pm \alpha_1\, , \\
B_\pm = \frac{1}{2} \pm \alpha_2\, .
\end{split}
\end{eqnarray}
Here $\alpha_1$ and $\alpha_2$ are arbitrary real constants.
In Ref. \cite{koide19}, we choose $(\alpha_1,\alpha_2) = (0,1/2)$ and 
quantum hydrodynamics in the curved spacetime is derived. 
See also the discussion in Refs. \cite{koide-review,koide18}.
Then we consider the variation of the stochastic action defined by 
\begin{eqnarray}
I = \int^{t_f}_{t_i} \ud t E[L_p]\, ,
\end{eqnarray}
where the stochastic Lagrangian is expressed by 
\begin{eqnarray}
&& L_p = \int \ud V_\xi \rho_{(0)} (\xi) \nonumber \\
 && \hspace*{-1cm}\times \left[
\frac{B_+}{2}  \sum_{i=\pm} A_i  (\uD_i \widehat{y}^{\, a})\eta_{ab} (\uD_i \widehat{y}^{\, b} )
+ \frac{B_-}{2} (\uD_+ \widehat{y}^{\, a}) \eta_{ab} (\uD_- \widehat{y}^{\, b})
- \frac{\varepsilon ( \rho_{(0)}(\xi) /J (\partial \widehat{x}_t) )}{ \rho_{(0)}(\xi) /J (\partial \widehat{x}_t)  } 
\right]\, . \label{eqn:sto-lag}
\end{eqnarray}
In the vanishing limit of $\nu$, $\uD_+$ and $\uD_-$ coincide with the standard time derivative and 
this Lagrangian is reduced to the ideal one (\ref{eqn:lag-euler-lag2}).
Note that  we cannot control the behavior of each stochastic motion because of the noise but 
optimize the expectation value of the action.

The variation of the stochastic trajectory is represented by 
\begin{eqnarray}
\widehat{x}^{\, i}_t \longrightarrow \widehat{x}^{\, i}_t  + \delta f^{i} (\widehat{x}_t)\, .
\end{eqnarray}
Here, by using the smooth infinitesimal function considered in Eqs.\ (\ref{eqn:smoothvari1}) and (\ref{eqn:smoothvari2}), 
the second term on the right-hand side is given by 
\begin{eqnarray}
\delta f^{i} (\widehat{x}_t) = \left. \delta f^{i} (\vec{x}_t,t) \right|_{{x}^{i}_t = \widehat{x}^{i}_t}\, .
\end{eqnarray}

The result of the stochastic variation is given by 
\begin{eqnarray}
\lefteqn{v^\mu \nabla_\mu v^{i} 
+ \frac{1}{\rho} g^{ij}\partial_j P} && \nonumber \\
&=&
 \kappa g^{jk}\left[  (\partial_k \ln \rho) \nabla_j  +  \nabla_j \nabla_k \right] g^{il} \partial_l \ln \rho 
- \xi \left[
v^\mu \nabla_\mu g^{ij} \partial_j \ln \rho - \frac{1}{\rho} g^{jk}\nabla_j (\rho \nabla_k v^i)
\right] \, , \label{eqn:opti-fluid}
\end{eqnarray}
where 
\begin{eqnarray}
\begin{split}
&\kappa = 2 \alpha_2 \nu^2\, ,\\
& \xi = 2\alpha_1 \left( \frac{1}{2} + \alpha_2\right) \nu\, .
\end{split}
\end{eqnarray}
In this derivation, we used 
\begin{align}
\begin{aligned}
& \delta \int^{t_f}_{t_i} \ud t E\left[ (\uD_+ \widehat{y}^{a})\eta_{ab}(\uD_+ \widehat{y}^{b}) \right] 
=
-2 \int^{t_f}_{t_i} \ud t E 
\left[
\delta f^j  \,  g_{jk} \left(  c\nabla_0 + u^l_-\nabla_l - \nu g^{lm}\nabla_l \nabla_m \right) u^k_+ 
\right]
\, ,\\
&\delta \int^{t_f}_{t_i} \ud t E\left[(\uD_- \widehat{y}^{a})\eta_{ab}(\uD_- \widehat{y}^{b}) \right]
=
-2 \int^{t_f}_{t_i} \ud t E 
\left[
\delta f^j \,  g_{jk} \left(  c\nabla_0 + u^l_+ \nabla_l + \nu g^{lm} \nabla_l \nabla_m \right) u^k_- 
\right]
\, ,\\
& \delta \int^{t_f}_{t_i} \ud t E\left[(\uD_+ \widehat{y}^{a})\eta_{ab}(\uD_- \widehat{y}^{b})\right]  \\
&  =
- \int^{t_f}_{t_i} \ud t E 
\left[
\delta f^j  \, g_{jk}
\left\{  \left(  c\nabla_0 + u^l_- \nabla_l - \nu g^{lm}\nabla_l \nabla_m \right) u^k_- 
+ 
 \left(  c\nabla_0 + u^l_+ \nabla_l + \nu g^{lm}\nabla_l \nabla_m \right) u^k_+ 
\right\}
\right]
\, . 
\end{aligned}
\end{align}
The stochastic variation of the internal energy term, which is independent of $\uD_\pm$,
gives the same result 
as the standard classical variation shown in Sec.\ \ref{sec:euler}.

To express Eq.\ (\ref{eqn:opti-fluid}) in a more familiar form, we notice that 
\begin{eqnarray}
\begin{split}
v^\mu \nabla_\mu g^{k \nu} \partial_\nu \ln \rho 
&=
-  \nabla_\nu  \nabla_\mu g^{k\nu} v^\mu - g^{k\nu}(\partial_\mu \ln \rho) \nabla_\nu v^\mu \, , \\
\protect{[\nabla_{\mu}, \nabla_{\nu}] A^{\alpha \beta \gamma}}
&=
R^\alpha_{\delta \mu \nu} A^{\delta \nu \gamma}
+ 
R^\beta_{\delta \mu \nu} A^{\alpha \delta  \gamma}
+
R^\gamma_{\delta \mu \nu} A^{\alpha \beta \delta }\, , 
\end{split}
\label{for:1}
\end{eqnarray}
where the Riemann curvature tensor is defined by 
\begin{eqnarray}
R^{\delta}_{\gamma \mu \nu} 
= \partial_\mu \Gamma^\delta_{\nu \gamma}
- \partial_\nu \Gamma^\delta_{\mu \gamma} 
+ \Gamma^\delta_{\mu \alpha}\Gamma^\alpha_{\nu \gamma}
- \Gamma^\delta_{\nu \alpha}\Gamma^\alpha_{\mu \gamma}\, .
\end{eqnarray}
The first equation of Eq.\ (\ref{for:1}) is obtained from the equation of continuity of the scalar mass distribution.
Then Eq.\ (\ref{eqn:opti-fluid}) is reexpressed as
\begin{eqnarray}
\rho v^\mu \nabla_\mu v^{i}+  g^{ij}\partial_j P  
= 
\nabla_\mu \{\xi \rho (g^{ik} \nabla_k v^\mu + g^{\mu k} \nabla_k v^{i}) \}
+ \xi \rho g^{ij}R_{j\alpha}v^\alpha + g^{ij} g^{kl}\nabla_l (\kappa \rho \nabla_j \partial_k \ln \rho)\, ,
\label{eqn:general-res} \nonumber \\
\end{eqnarray}
where the Ricci tensor is defined by $R_{\mu\nu} = \sum_{\alpha =0}^D R^\alpha_{\mu\nu\alpha}$.
This is our hydrodynamical equation in the curved spacetime and the principal result of this paper.
 Because of the time dependence of the geometry, 
the first term on the right-hand side contains a time derivative term.

In Ref.\ \cite{koide19}, it is reported that the interplay between quantum fluctuation and spacetime curvature induces the quantum curvature term which 
works in the same direction as the effects of the dark energy and the dark matter in cosmology.
The corresponding term exists even in the present calculation but is hidden in the last term on the right-hand side of the above equation.
See Eq.\ (24) of Ref.\ \cite{koide19}.

\section{Specific limits} \label{sec:incom}

In this section, we study some special cases of the general form of Eq.\ (\ref{eqn:general-res}).

\subsection{Static geometry limit with vanishing $\kappa$}

The last term on the right-hand side of Eq.\ (\ref{eqn:general-res}) corresponds to the possible modification of the NSF equation proposed by Brenner, 
which is discussed later.
To ignore this contribution for a while, we consider the case of $\kappa= 0$ (or equivalently $\alpha_2 =0$).
For a further simplification, we consider the  static spacetime geometry, $\partial_0 g_{\mu\nu}=0$ and then we can show 
that the time derivative term in the first term on the right-hand side vanishes
\begin{eqnarray}
\nabla_\mu \rho g^{i\nu}\nabla_\nu v^\mu + \rho R_{\nu \mu} g^{i\mu} v^\nu 
= 
\nabla_j \rho g^{ik}\nabla_k v^j + \rho R_{k j} g^{ij} v^k\, .
\end{eqnarray}

With this, Eq.\ (\ref{eqn:general-res}) reduces to
\begin{eqnarray}
\rho (\partial_t  + v^{j} \nabla_j )  v^{i}+  g^{ij}\partial_j (P - \zeta \theta ) = 
 \nabla_j \eta E^{ij}  +  \frac{\eta}{2} g^{ij} R_{j k}v^k \, , \label{eqn:pot-nsf} 
\end{eqnarray}
where 
\begin{eqnarray}
\begin{split}
& \theta = \nabla_j v^j\, , \\
& E^{ij} = \frac{1}{2} \left( g^{ik}\nabla_k v^{j} + g^{jk}\nabla_k v^{i} \right) - \frac{1}{D} \theta g^{ij} \, .
\end{split}
\end{eqnarray}
The coefficient of viscosity and the bulk viscosity are defined by 
\begin{eqnarray}
\begin{split}
& \eta = 2 \xi \rho\, ,\\
& \zeta = \frac{ \eta}{D}\, ,
\end{split}
\end{eqnarray}
respectively.
Note however that the contribution from the second coefficient of viscosity is not considered in the present calculation. 
See also the discussion in Sec.\ \ref{sec:con}.
This equation is the compressible hydrodynamical equation on static spatial Riemannian manifolds.
The first term on the right-hand side corresponds to the contributions from the shear viscosity. 
The correction to the adiabatic pressure appears as the bulk viscosity in the second term on the left-hand side.
The second term on the right-hand side is induced by the interplay between thermal fluctuation and spatial curvature. 
One can easily see that Eq.\ (\ref{eqn:pot-nsf}) reproduces the standard NSF equation in the Euclidean limit.

Equation (\ref{eqn:pot-nsf}) can be cast into the form of 
the equation of continuity,
\begin{eqnarray}
\partial_t (\rho  v^{i}) = - \nabla_j T^{ij}\, ,
\end{eqnarray}
where the fluid-stress tensor is defined by 
\begin{eqnarray}
T^{ij} 
&=&  \rho v^i v^j  + \left\{ P -  \nabla_k \left( \frac{\eta}{2} v^{k} \right) -  \zeta \theta   \right\} g^{ij} 
- \eta E^{ij}
+ g^{ik} \nabla_k \left( \frac{\eta}{2} v^{j} \right) 
%  -\kappa \rho g^{ik}g^{jl} \nabla_{l}\partial_j \ln \rho 
 \, . \label{eqn:tij}
\end{eqnarray}
The contribution from the Ricci tensor is represented by the third and last terms on the right-hand side.
Note that this tensor is not symmetric for the exchange of the indices $i$ and $j$, 
differently from that of the Euclidean NSF equation. 
This means that the symmetry of the fluid-stress tensor in the Euclidean NSF equation is not valid for hydrodynamics in the curved systems.
It is also worth reminding that such a violation is observed also by the 
internal degrees of freedom of fluid \cite{koidespinhydro}.
See also discussion in Sec.\ 2 of Ref. \cite{chan}.

Let us further consider the incompressible limit, 
$\nabla_i v^i = 0$.
Then Eq.\ (\ref{eqn:pot-nsf}) is reduced to 
\begin{eqnarray}
(\partial_t  + v^{j} \nabla_j ) v^{i}+  \frac{g^{ij}}{\rho}\partial_j P  = 
\frac{\eta}{2\rho} g^{jk}\nabla_j \nabla_k v^{i}\, .\label{eqn:incom-nsf}
\end{eqnarray}
where $\rho$ is a constant here.
This is our incompressible hydrodynamical equation.
As was mentioned in the introduction, there are three different proposals 
in the literature to replace the Laplacian in the viscous term which corresponds to the right-hand side of the above equation. 
In our formulation,  
the corresponding Laplacian 
is given by 
the Bochner Laplacian.

\subsection{Role of $\kappa$ term}

Our most general result given by Eq.\ (\ref{eqn:general-res}) contains an additional contribution to the NSF equation, which depends on the parameter $\kappa$.
There are two different interpretations to understand the role of the $\kappa$ term.
In quantum mechanics, it is known that the  Schr\"{o}dinger equation can be expressed in the form of hydrodynamics by replacing the pressure with 
the so-called quantum potential \cite{book:holland,koide19}. 
The $\kappa$ term formally corresponds to this \cite{koide12,koide18}, 
but we consider here different origins: the quantum potential term is caused by quantum fluctuations while the $\kappa$ term is induced by thermal fluctuations.

Another interpretations is the possible modification of the Euclidean NSF equation proposed by Brenner 
\cite{brenner1,brenner2,brenner3,klimontovich,ottinger,graur,greensh,eu,don,dadzie,gustavo}. 
He pointed out that, since the velocity of a tracer particle of fluids is not necessarily parallel to the mass velocity ($v^{i}$ in our notation),
the existence of these two velocities should be taken into account in the formulation of hydrodynamics. 
In this context, by applying the linear irreversible thermodynamics, the difference of the two velocities is found to be characterized 
by the mass density gradient as is given by our consistency condition (\ref{eqn:cons-cond}) and a new effect corresponding to the $\kappa$ term appears. 
See also the Table 1 in Ref.\ \cite{brenner2}.
This theory is called bivelocity hydrodynamics but the introduction of such a modification is still controversial. 
It is also worth mentioning that the structure analogous to bivelocity hydrodynamics 
naturally appears as the next-to-leading order relativistic corrections to the NSF equation \cite{gustavo}.

\section{Comparison with other SVM calculations} \label{sec:compa}

The different formulations of SVM are developed in Refs.\ \cite{yasue-ns,cruzeiro} for the incompressible fluid on spatial Riemannian manifolds.

Differently from ours, in Ref.\ \cite{yasue-ns}, the mean forward and backward derivatives are defined by considering  
the effect of the parallel transport of vector. 
However, the applicability of the stochastic partial integration formula (\ref{eqn:pif}) is not clear for such a modification.
On the other hand, in our formulation, the standard definitions used in the Euclidean formulation are still employed while 
the derivatives operated to tetrads give the corrections corresponding to the parallel transport. See the second line of Eq.\ (\ref{eqn:ito-vector2}).
Therefore Eq.\ (\ref{eqn:pif}) is applicable.
Despite these differences, the result of Ref.\ \cite{yasue-ns} is equivalent to our incompressible result  (\ref{eqn:incom-nsf}) and 
the Laplacian in the viscous term is given by the Bochner Laplacian.

Another approach in Ref.\ \cite{cruzeiro} gives a different result and the corresponding Laplacian is given by the 
Hodge Laplacian. 
To reproduce the same result in our formulation, 
we should, for example, change the definition of the stochastic parallel transport, as is considered in Refs.\ \cite{dunkel,dohrn1,dohrn2}.
See, for example, Eq.\ (173) in Ref. \cite{zam-review}.
Then the Bochner Laplacian in Eq.\ (\ref{eqn:ito-vector2}) gains an additional contribution associated with the Ricci tensor. 
See Eq.\ (180) in Ref. \cite{zam-review}.
Such a parallel transport, however, does not conserve the length of vector \cite{zam-review}.
Therefore we do not consider the different stochastic parallel transport in this paper.

\section{Gauge interaction} \label{sec:gauge}

One of the advantages in variational approaches is its compatibility with the symmetry principle. 
As one of such applications, we consider the gauge interaction of a charged compressible fluid 
using the gauge-invariant Lagrangian and show that the Lorentz force appears as the interaction between the Abelian gauge field and viscous charged fluid. 
For simplicity, we consider the static geometry.

Introducing the electric charge per unit mass $q$, the 
conserved charge current is defined by 
\begin{eqnarray}
J^\mu (x)=\frac{ q} {c}\rho(x) v^\mu (x)\, ,
\end{eqnarray}
satisfying $\nabla_\mu J^\mu = 0$.
Then the gauge-invariant stochastic Lagrangian is defined by 
\begin{eqnarray}
L &=& L_p + \int \ud V_\xi \rho_{(0)}(\xi) \left[
 \frac{q}{c} \eta_{ab} \frac{\uD_+ \widehat{y}^{\, a}_t + \uD_- \widehat{y}^{\, a}_t}{2} \overline{e}^b_\nu (\widehat{x}_t) A^\nu (\widehat{x}_t)
\right] \, ,
\end{eqnarray}
where $A^\mu (x)$ is the Abelian gauge field and $L_p$ is defined by Eq.\ (\ref{eqn:sto-lag}).

Substituting into the stochastic action and calculating the stochastic variation, we find  
\begin{eqnarray}
\rho v^\mu \nabla_\mu v^{i}+  g^{ij}\partial_j (P - \zeta \theta ) = 
 \nabla_j \eta E^{ij}  +  \frac{\eta}{2} g^{ij} R_{j \alpha}v^\alpha
+ g^{ij}g^{kl} \nabla_{l}( \kappa \nabla_j \partial_k \ln \rho)
+ g_{jk}J^j f^{ik}
\, , 
\end{eqnarray}
where the field-strength tensor is defined by 
\begin{eqnarray}
f^{\mu\nu} \equiv g^{\mu\alpha}g^{\nu\beta} f_{\alpha \beta}\, , \, \, \, \, 
f_{\mu\nu} = \partial_\mu (g_{\nu \alpha}A^\alpha) - \partial_\nu (g_{\mu\alpha} A^\alpha) \, .
\end{eqnarray}
The last term on the right-hand side of the derived equation is the well-known Lorentz force.

\section{concluding remarks} \label{sec:con}

In this paper, we derived the non-relativistic hydrodynamical equation for a compressible fluid in
the curved spacetime, using the generalized stochastic variational method
for continuum medium developed in Refs.\ \cite{koide12,koide19}. The
principal result is given by Eq.\ (\ref{eqn:general-res}). 
For the sake of comparison, we considered the incompressible limit of Eq.\ (%
\ref{eqn:general-res}) with the static geometry and found that our
variational approach sustains the use of the Bochner Laplacian to express
the viscous term. We further found that, even in curved geometries, it is still possible to consider Brenner's modification of the NSF equation 
which has the similar form to the quantum potential \cite{koide19}. To show the compatibility with the symmetry principle, the
gauge-invariant Lagrangian of a charged compressible fluid was considered
and the Lorentz force was reproduced as the interaction between the Abelian
gauge field and the viscous charged fluid.

There are two other different formulations of SVM which are applicable to the
incompressible fluid on static spatial Riemannian manifolds \cite{yasue-ns,cruzeiro}. 
The incompressible limit of our result coincides with the result of Ref.\ 
\cite{yasue-ns}, but is different from that of Ref.\ \cite{cruzeiro}. 
To reproduce Ref.\ \cite{cruzeiro} in our formulation, 
the definition of the stochastic parallel transport should be changed, but then the
length of the transported vector is not conserved.

In the derivation of the Euclidean NSF equation, the fluid-stress tensor is
normally assumed to be symmetric for the exchange of the indices, 
but such a symmetry is not observed in our result.
Hence the standard structure familiar to the Euclidean NSF
equation is not necessarily found for hydrodynamics in the curved spacetime.
It is then interesting to consider the similar symmetry appearing in the
energy-momentum tensor of relativistic fluids. Although the structure is rather
modified from the NSF equation due to the relativistic kinematics 
\cite{hydroreview}, the symmetry still plays a critical role in (the derivation
of) relativistic hydrodynamics (so-called Israel-Stewart-type theory).
Therefore the above mentioned violation of the symmetry will have a profound
effect in the relativistic fluid under the general relativity, such as the
stability analysis \cite{nor}.

It is worth mentioning that the formulation of hydrodynamics based on SVM
meets our understanding for viscosity: viscosity is caused by
interactions among fluid elements. In SVM, such an interaction is introduced
through thermal fluctuations (noise) in the SDE's, leading to Brownian motion of the fluid element. 
If we observe the fluid in a macroscopic
scale where the zigzag of the trajectory is negligible, we can apply the
standard classical variation to the fluid Lagrangian 
and obtained is the Euler equation. On the other hand,
when we observe the same system with a relatively smaller scale and 
the effect of the zigzag is considered, the stochastic variation should be applied to the
same Lagrangian under an appropriate stochastic replacement of variables and then the viscous term appears as the influence of the
fluctuation of the fluid elements. Of course, then the Euler equation is
reproduced as the vanishing limit of thermal fluctuations in the NSF
equation.

In this paper, the finite contribution of the second coefficient of
viscosity is not considered and then the bulk viscosity $\zeta$ is
given by $\eta/D$. To reproduce this coefficient, for example, we can
consider the variation of the entropy dependence in the pressure $P$, as is
discussed in Ref.\ \cite{koide12}.
Moreover we have considered the barotropic fluid where the
thermodynamic pressure $P$ is adiabatic and given by a function only of the
scalar mass distribution and thus the set of Eqs.\ (\ref{eqn:eoc}) and 
(\ref{eqn:general-res}) is already closed. However, when the entropy dependence is
included in $P$, one more equation from the energy conservation is
necessary. The energy equation of the inviscid relativistic fluid is derived from the
Noether theorem. See, for example, \cite{hugo}. It is still an open question
whether the similar argument is applicable to reproduce the viscous energy
equation in SVM.

As was mentioned in the introduction, the hydrodynamical model 
is considered to be successful in the studies of the macroscopic behaviors of the quark-gluon plasma (QGP)
produced in nucleus-nucleus (A-A) collisions 
through the collective behaviors of relativistic particles \cite{hydroreview}. 
It is normally assumed that quantum effects affect 
only the parameters of the model (equation of state, coefficients of viscosity, etc.), 
and the classical equations are simply employed. 
However the quantum effect in dynamics should be important   
to understand the emergence  of the collective flow in smaller systems such as proton-proton (p-p), He-He and proton-nucleus (p-A) collisions. 
Moreover, by the recent advances in the observations of gravitational waves,  
binary neutron star mergers offer an opportunity to study highly dense quark-hadronic matter, 
with a view to data obtained from heavy-ion collisions. 
See, for example,  \cite{stocker1,stocker2,stocker3}.  
The present approach may provide a procedure to find a modified hydrodynamics to describe these situations.

As is shown in Refs.\ \cite{koide18,ucr-koide}, hydrodynamics and quantum mechanics
are formulated on an equal footing in SVM and thus it is possible to define
the uncertainty relations even for fluid as the natural generalization of the
well-known quantum mechanical ones; the Kennard inequality and the
Robertson-Schr\"{o}dinger inequality. The same argument will be applicable
to curved spacetime systems. 
The uncertainty relations in hydrodynamics will represent a certain aspect of the QGP fluid  
for energy distributions in the presence of spiky inhomogeneities.
These are left as future tasks.

\vspace{2cm}

T.\ Koide thanks J.-P.\ Gazeau and C.\ A.\ D.\ Zarro for useful discussions and comments. 
The authors acknowledge the financial support by CNPq(303468/2018-1), FAPERJ, CAPES and PRONEX. 
A part of the work was developed under the project INCT-FNA Proc. No. 464898/2014-5.

\appendix

\section{Forward SDE in Ito definition} \label{app:ito}

The SDE's can be reexpressed by substituting the Stratonovich definition (\ref{eqn:stra}).
The forward equations  are expressed as
\begin{eqnarray}
\begin{array}{l}
\ud \widehat{x}^{\,i}_t  
= 
\tilde{u}^i_+ \ud t
+ \sqrt{2\nu} \underline{e}^i_a   \ud \widehat{W}^a_t \, , \\
\ud \underline{e}^\mu_a 
= 
\left(
-\Gamma^\mu_{\gamma \delta} \tilde{u}^{\gamma}_+ 
- \nu g^{kj} \partial_j \Gamma^{\mu}_{k\delta} 
+ \nu \Gamma^{\mu}_{k\nu}\Gamma^{\nu}_{j\delta} g^{kj}
\right) \underline{e}^\delta_a \ud t
- \sqrt{2\nu}\Gamma^\mu_{j\delta} \underline{e}^\delta_a \underline{e}^j_b \ud \widehat{W}^b_t\, ,
\end{array}
\end{eqnarray}
where, for the simplicity of notation, we introduced $\tilde{u}^\mu_+ = (c, \tilde{u}^{i}_+)$ with 
\begin{eqnarray}
\tilde{u}^{i}
= 
u^i_+ -\nu \Gamma^{i}_{jk} g^{jk}\, .
\end{eqnarray}
The corresponding equation for the backward SDE can be obtained in the same way.

\section{Periodic variable} \label{app:periodic}

The present definition of the scalar mass distribution $\rho$ should be modified when we choose a periodic variable as a component of generalized coordinates.
For example, let us consider polar coordinates described by $(r, \theta)$.
The evolution of the angle coordinate $\theta$ is described by the SDE's and then $\widehat{\theta}_t$ can take any real value, 
$-\infty < \widehat{\theta}_t < \infty$. Note however that the unknown functions in the SDE's are periodic, 
\begin{eqnarray}
u^i_\pm (r, \theta, t) = u^i_\pm (r, \theta+2\pi, t) \, .
\end{eqnarray}

Then we can define the scalar mass distribution for the bounded domain of the angle parameter in Eulerian coordinates $0\le \theta < 2\pi$ as 
\begin{eqnarray}
\rho(r, \theta,t) = \frac{1}{\sqrt{-g}} \int \ud V_\xi \, \rho_{(0)}(\xi) 
 \sum_{N=-\infty}^\infty E[\delta (r - \widehat{r}_t (\xi))\delta (\theta + 2\pi N - \widehat{\theta}_t (\xi))]\, ,
\end{eqnarray}
where the sum runs over all integers.
Because of the periodicity for $u^i_\pm (r, \theta, t)$, this $\rho$ still satisfies the same Fokker-Planck equations (\ref{eqn:fpf}) and (\ref{eqn:fpb}) 
and thus the present result of SVM is not changed even for the periodic variables. 
See also the discussion in Ref.\ \cite{ucr-koide}.

\end{document}